\documentclass[a4paper,11pt]{article}
\usepackage{pos}
\usepackage{graphicx}
\usepackage{mathtools}
\usepackage{amsmath,amsfonts,amsthm}
\usepackage{array}
\usepackage{multirow}
\usepackage{url}
\usepackage{aas_macros}

\title{Search for cosmic-ray induced gamma-ray emission from local galaxy clusters using {\it Fermi}-LAT data }
\ShortTitle{Fermi-LAT - Clusters diffuse $\gamma$-ray emission}

\author*[a]{Judit Pérez-Romero}
\author[b,c]{Mattia di Mauro}
\author[d]{Rémi Adam}
\author[e]{Miguel Á. Sánchez-Conde}
\author[a]{Gabrijela Zaharijas}

\affiliation[a]{Center for Astrophysics and Cosmology, University of Nova Gorica, Vipavska 11c, 5270 Ajdovščina, Slovenia}
\affiliation[b]{Dipartimento di Fisica, Universit\'a di Torino, Via P. Giuria 1, 10125 Torino, Italy}
\affiliation[c]{Istituto Nazionale di Fisica Nucleare, Sezione di Torino, Via P. Giuria 1, 10125 Torino, Italy}
\affiliation[d]{Université Côte d'Azur, Observatoire de la Côte d'Azur, CNRS, Laboratoire Lagrange, France}
\affiliation[e]{Instituto de F\'isica Te\'orica, IFT UAM-CSIC, Departamento de F\'isica Te\'orica, Universidad Aut\'onoma de Madrid, ES-28049 Madrid, Spain}

\emailAdd{judit.perez@ung.si}

\abstract{Galaxy clusters are the most massive gravitationally bound structures in the Universe. Even if clusters are nearly virialized structures, they undergo merging processes, creating merging shocks, and suffer from feedback from galaxies and Active Galactic Nuclei; causing complex turbulent motions and amplifying their magnetic fields. These processes act as acceleration mechanisms for the plasma of the intracluster medium (ICM), originating a population of cosmic rays (CRs). Leptonic CRs have long been detected, but we should also expect a CR hadronic population that, through interactions with the ICM, should produce neutral pions that decay into $\gamma$-rays. The detection of diffuse $\gamma$-ray emission from galaxy clusters is one of the long-awaited milestones for the high-energy astroparticle physics community. Still, no unambiguous detection has yet been obtained.\\ 
In this talk, we will present the results of a combined cluster analysis searching for CR-induced $\gamma$-ray signals, using 16 years of {\it Fermi}-LAT data. In our previous work (di Mauro et al. 2023) we obtained from the combined analysis of 49 local galaxy clusters (12 years of data) a hint of signal between 2.5-3$\sigma$. %, depending on the DM model considered. 
These results are consistent with other works as well, %are aligned with the most recent works on searches for $\gamma$-ray emission from clusters on {\it Fermi}-LAT data, 
which consistently find a non-vanishing hint of signal, around the detection threshold. In this new work, we use a sample of near, well-known galaxy clusters and develop CR-induced emission templates using well-established X-ray measurements for calibration, assuming self similarity for the members of our sample. To strengthen the robustness of our analysis, we define benchmark models to encapsulate the uncertainties in the spectral and spatial profiles for the CR-induced emission and perform the standard template-fitting analysis using the likelihood ratio test.}

\ConferenceLogo{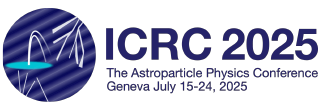}

\FullConference{39th International Cosmic Ray Conference (ICRC2025)\\
 15–24 July 2025\\
Geneva, Switzerland\\}
\begin{document}
\maketitle

\section{Introduction}
According to the standard cosmological model, structures in the Universe grow following a bottom-up hierarchical scenario, where the smaller structures form first. Galaxy clusters are the largest gravitationally bound systems in the Universe, and their physical properties are outlined by their formation process. The gravitational energy from smooth accretion and merging of subclusters is mostly dissipated into the thermal, ionized gas called the intracluster medium (ICM). These shock waves from the formation processes propagating in the ICM and turbulence are expected to accelerate both electrons and protons at relativistic energies, producing a non-thermal population of cosmic rays (CRs) confined within the cluster magnetic fields (see \cite{2014IJMPD..2330007B} for a review). Despite galaxy clusters are usually assumed as stable, virialized objects, other astrophysical processes can also contribute to the creation of a CR population, such as active galactic nuclei (AGN) outbursts (e.g. \cite{Bonafede:2014zfa}) and galactic winds from star formation activity in member galaxies (e.g. \cite{Rephaeli:2016yhx}). 

In fact, radio observations of diffuse synchrotron emission in several clusters \cite{Botteon:2022dxk} have confirmed the presence of CRe and magnetic fields. This emission can be classified between radio halos and relics, depending on their spatial properties and their hydrodynamical state. CR protons (CRp) should also be retained in galaxy clusters because of their long lifetimes. These CRp should interact hadronically in the ICM to produce $\gamma$-ray emission as a product of $\pi^{0}$ decays \cite{Hussain:2022tls}. Such interactions also lead to the production of secondary CRe, which will contribute to the observed radio emission. 

The search for diffuse $\gamma$-ray emission from galaxy clusters has been going on for more than two decades, and despite all efforts. % from the astroparticle physics community, such a signal still remains elusive. 
The null detection has been translated into constraints on the CRp-to-thermal energy ratio, and the spectral and spatial profiles of the CRp population in clusters. However, investigating the most recent literature, we can notice a moderate trend to find a mild detection ($\sim 2-5\sigma$, see e.g. \cite{Ackermann:2013iaq, AdamEtAl2021, Manna:2025fvf}). In particular, in our previous work \cite{DiMauro:2023qat}, we used 12 years of {\it Fermi}-LAT data to search for DM-induced $\gamma$-ray emission in the direction of 49 local clusters, resulting in $TS = 25-30$ that translates into a $2.5-3.0\sigma$ detection, after applying corrections from using blank sky directions for the null hypothesis.

In this new work, we want to further explore the $2.5-3.0\sigma$ hint from \cite{DiMauro:2023qat} within the framework CR-induced $\gamma$-ray emission. We focus on the same cluster sample but with an increased amount of {\it Fermi}-LAT data, using the 16 years available at the time of starting the analysis.

\section{Modelling the CR-induced $\gamma$-ray emission in clusters}
\label{section:modelling}

The analysis sample includes 48 local galaxy clusters, all of them located at $|b|>20$ deg, to avoid any possible contamination from Galactic diffuse emission, and selecting only those at $z< 0.1$ to prevent significant attenuation. For a more detailed description of the sample, see Section~II of our previous work \cite{DiMauro:2023qat}. %The main updates we did on the sample are:
%\begin{enumerate}
%    \item We removed galaxy cluster M49, since it is merging with Virgo cluster, thus overlapping when considering their extensions;
%    \item We updated the estimations for the masses $M_{500}$ to the ones from the MCXC catalogue \cite{2011A&A...534A.109P}, that shows better agreement with masses from Planck SZ cluster catalogue \cite{2016A&A...594A..27P}.
%\end{enumerate}

CRs are predicted to generate photons across the entire electromagnetic spectrum via synchrotron radiation, % in the intracluster magnetic fields, 
bremsstrahlung and $\pi^{0}$ decays through the interaction with the ICM, and inverse Compton (IC) scattering on the photon fields. The dominant production mechanism for $\gamma$-rays in the energy range from 10 MeV - 1 TeV, operating range of {\it Fermi}-LAT, is the decay of neutral pions $\pi^{0}$ (see Figure~2 in \cite{CTAConsortium:2023yak}):
\begin{equation}\label{eq.1}
    {\rm CRp} + {\rm p_{ICM}} \rightarrow \pi^0 \rightarrow \gamma\gamma.
\end{equation}

The $\gamma$-ray emission induced by hadronic interactions is directly related to the spatial and spectral distribution of CRp in the ICM, but also to the thermal gas on which they interact. To compute the $\gamma$-ray emission, we follow the approach described in \cite{Adam:2020atc}, which introduces the open-source software \texttt{MINOT}. The \texttt{MINOT} code aims at predicting ICM observables, including those in the gamma-ray band, based on the modeling of the relevant physical properties under the assumption of spherical symmetry. %code assumes that clusters are spherically symmetric and in hydrostatic equilibrium, but we expand this assumption by adding second-order corrections based on the hydrodynamical state of each cluster. To model the $\gamma$-ray diffuse emission of the 48 clusters and perform the stacking analysis with {\it Fermi}-LAT data, we assume that clusters are self-similar objects \cite{2012A&A...539A.120B}. With these assumptions, we can build $\gamma$-ray emission models for the clusters by parametrizing the proton number density $n_{p}$, the thermal gas pressure profile $P_{\rm gas}$ and the CRp distribution $\frac{dN_{\rm CRp}}{dEdV}$. 
The thermal electron number density and pressure profiles, which are required inputs, are modelled using the universal density and pressure profile parametrization from \cite{2022A&A...665A..24P} and \cite{2013A&A...550A.131P}, respectively, given the mass and the redshift of each cluster assuming self-similarity. This provides a good first order estimate of the thermodynamical profiles for each cluster. We use the ACCEPT data \citep{2009ApJS..182...12C} to correct the cluster core (normalization and inner slope) and better describe the observations, whenever deviation from the original model may be significant. For the Coma cluster, which is one of the 4/48 systems for which ACCEPT data are not available, we use the density and pressure profile given in \cite{AdamEtAl2021}.

\begin{figure}[h!]
\centering
\includegraphics[width=0.5\linewidth]{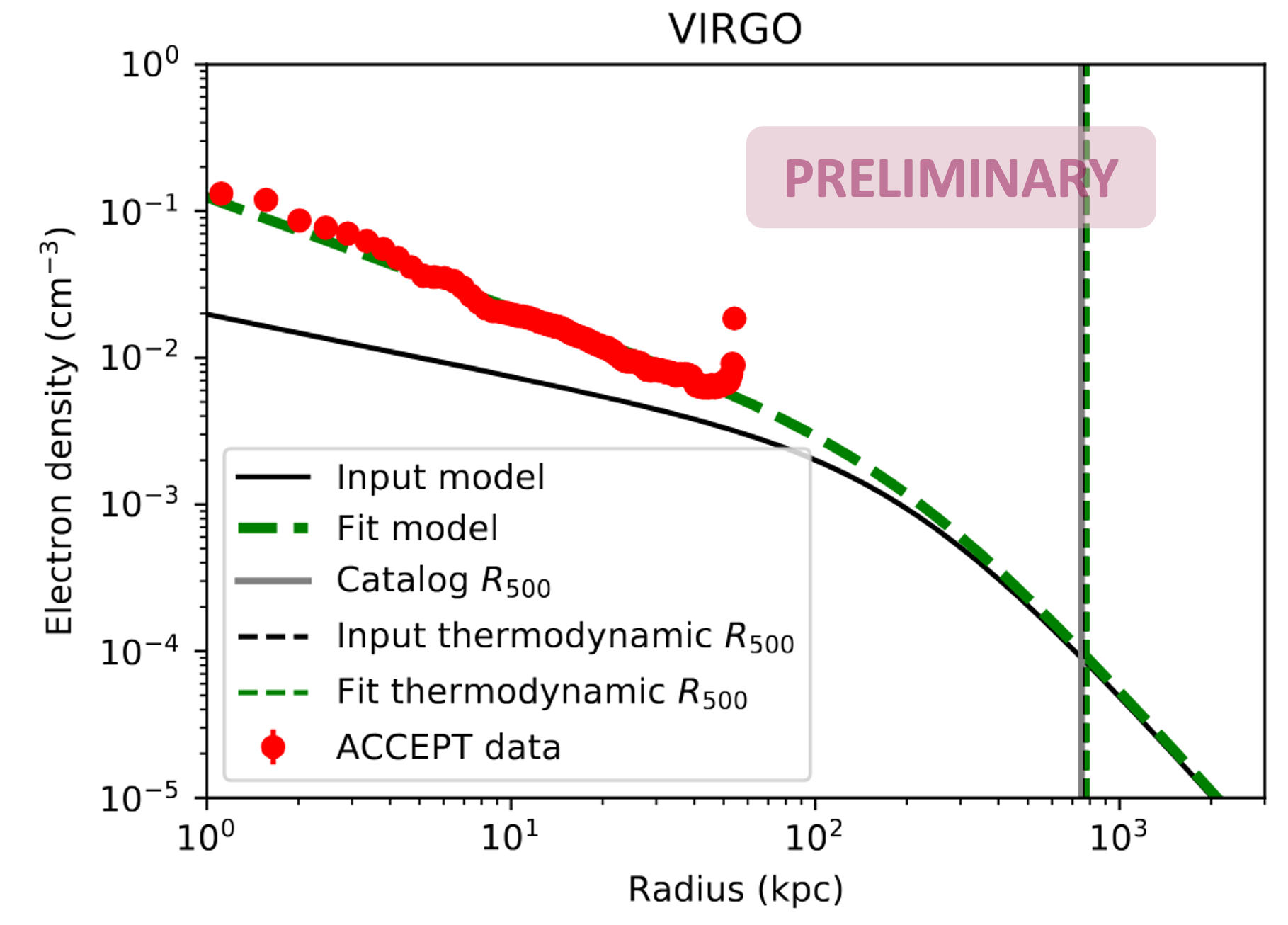}
\includegraphics[width=0.49\linewidth]{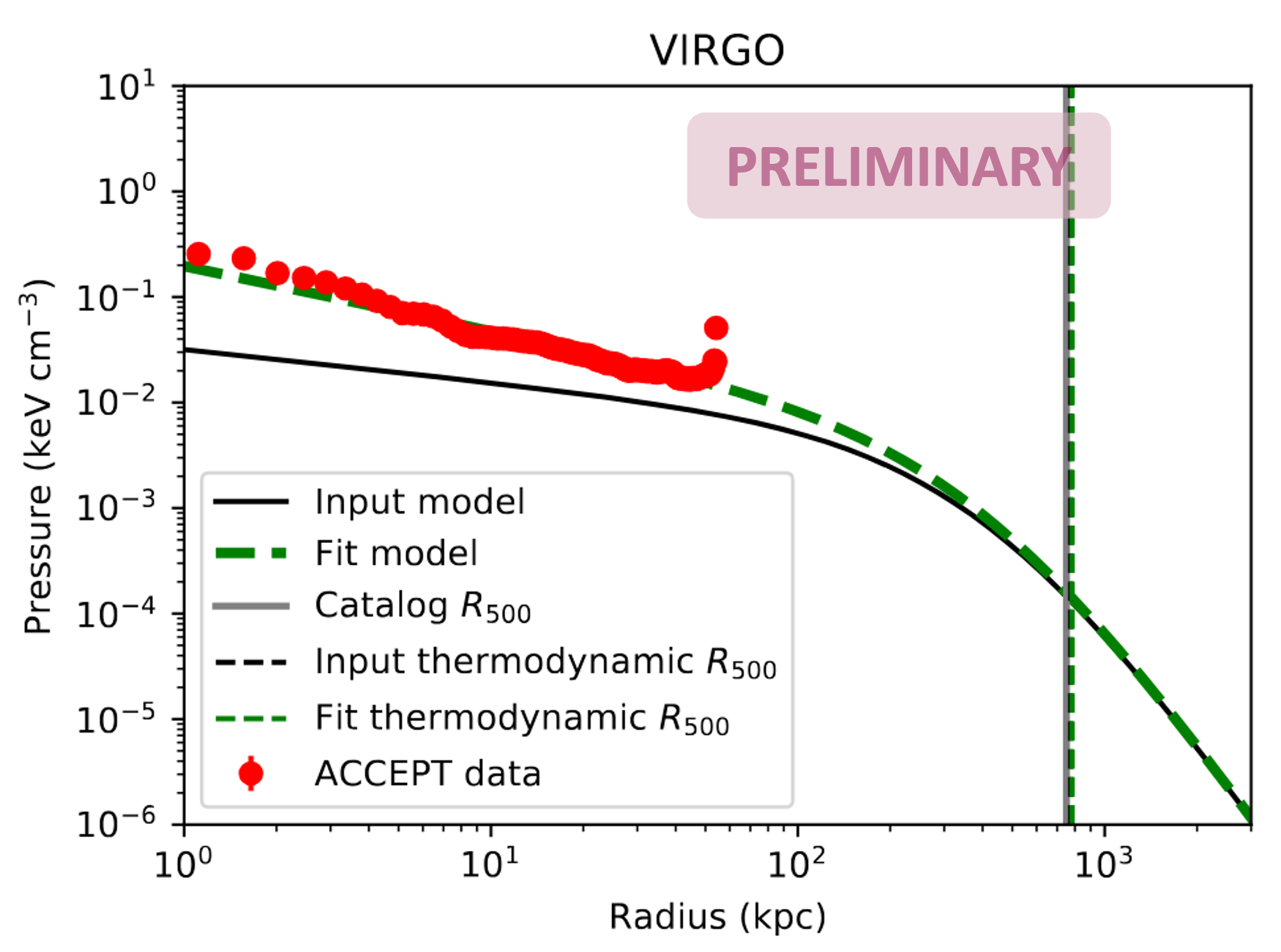}
\caption{Electron number density (left panel) and electron pressure (right panel) profiles obtained for the Virgo cluster using the parametrized models from \cite{2022A&A...665A..24P} and \cite{2013A&A...550A.131P}, respectively (solid black line). These parametrizations are compared to ACCEPT data \cite{2009ApJS..182...12C} (red dots). For most of the clusters in our sample with ACCEPT data available, the generalized models yield flatter profiles compared to the fitted profiles (dashed green line), being the former more conservative in terms of $\gamma$-ray emission.}
\label{fig.1}
\end{figure}

The thermal proton number density $n_{p}$ is obtained from the electron number density $n_{e}$, while the thermal gas pressure profile $P_{\rm gas}$ is derived from the electron pressure profile $P_{e}$, from the relations:
\begin{equation}\label{eq.2}
    n_{p} = \frac{\mu_e}{\mu_p} n_{e}, \;\;\;\;\;\;\; P_{\rm gas} = \frac{\mu_e}{\mu_{\rm gas}} P_{e},
\end{equation}
where $\mu_{X}$ represents the molecular weight of each component, respectively. %The electron number density $n_{e}$ can be obtained from X-ray observations for each individual cluster, same as the electron pressure profile $P_{e}$ can be proved by thermal SZ measurements. From the analysis of X-ray and thermal SZ data in a large sample of clusters, generalized parametric profiles can be derived, based on the aforementioned self-similarity of clusters, for both quantities. In this work, we follow the universal profiles from \cite{Ghirardini:2018byi} for $n_{e}$ and from \cite{2010A&A...517A..92A} for $P_{e}$, which are validated using ACCEPT data \cite{2009ApJS..182...12C} whenever available. 
An example of the modelling of $n_{e}$ and $P_{e}$ for the case of the Virgo cluster is shown in Figure~\ref{fig.1}. 

Finally, the CRp distribution is modeled according to a radial profile and
a canonical power-law:
\begin{equation}\label{eq.4}
    \frac{dN_{\rm CR_{p}}}{dEdV}(E, r)=A_{\rm CR_{p}}\times E^{-\alpha_{\rm CR_{p}}}\times n_{ p}^{\eta_{\rm CR_{p}}}(r).
\end{equation}
The radial profile assumes a scaling with respect to $n_{p}$ and is described by $\eta_{\rm CR_{p}}$, which accounts for the competition between advection and streaming. The slope of the power-law, $\alpha_{\rm CR_{p}}$, is related to the acceleration of CRs. The normalization $A_{\rm CR_{p}}$ is computed for each cluster given a value of the CRp-to-thermal energy ratio, $X = \frac{U_{\rm CR_{p}}}{U_{\rm th}}$, by integrating Equation~\ref{eq.4} up to $R_{500}$. 

The main modelling parameters of the CRp distribution, $X_{500}$, $\alpha_{\rm CR_{p}}$ and $\eta_{\rm CR_{p}}$ can be estimated using numerical simulations (e.g. \cite{Pinzke_2010, Bykov:2019bhd}), as well as derived from constraints from the lack of detection of diffuse $\gamma$-rays (e.g. \cite{MAGIC:2016jpr}). The current allowed parameter space yields between $\alpha_{\rm CR_{p}}\in[2, 3]$, $\eta_{\rm CR_{p}}\in[0, 1]$ and $X_{500}\sim 10^{-2}$. Uncertainty in these parameters can change the expected $\gamma$-ray fluxes by more than an order of magnitude. In the {\it Fermi}-LAT energy range, the $\gamma$-ray flux decreases with increasing $\alpha_{\rm CR_{p}}$ and the profile becomes more compact with increasing $\eta_{\rm CR_{p}}$. In order to address this intrinsic uncertainty, we define the following benchmark models that represent the different possible scenarios: 
$$X_{500}=1\%, \alpha_{\rm CR_{p}} = 2.0, 2.3, 2.6, 2.9;\;\; \eta_{\rm CR_{p}} = 1.0.$$ 
The radial distribution of CRs is expected to roughly scale with the thermal gas density ($\eta_{\rm CR_{p}} = 1$) when advection by the turbulent gas dominates, and since it is degenerated with $X_{500}$, the main parameter that we will fit during the analysis, we decided to keep it constant. Still, the spatial profile may flatten if diffusion and streaming become significant. Our baseline model, for which our main results will be presented unless stated otherwise, is then defined for: $$X_{500}=1\%,\;\alpha_{\rm CR_{p}} = 2.3,\; \eta_{\rm CR_{p}} = 0.75.$$

The final output of our CR modelling is a set of 48 three-dimensional templates containing the spatial morphology and the spectral information of each benchmark model, summing in total $48\times 5 = 240$ templates. These templates are obtained using the \texttt{MINOT} software. 

The integration of these templates on the energy range from 500 MeV - 1 TeV provides an estimate of which clusters should yield higher $\gamma$-ray fluxes, thus, which ones should dominate the {\it Fermi}-LAT analysis. Theoretically, the top 10 brightest clusters are, in descending order: A1656 (Coma), A3526 (Centaurus), A3571, A2199, 2A0335, A1367, A0085, A0496, A2029, A1795.

\section{{\it Fermi}-LAT analysis}
\label{sec:analysis}

The main choices for the analysis parameters are summarized in Table~\ref{tab.1}. We decide to neglect the lower-end of the energy range  since it is dominated by astrophysical background emission, having both a poor angular and energy resolution, which becomes even more acute for the case of very extended sources as galaxy clusters. Along these lines, the modelling of the Galactic interstellar emission (IEM) is also key. We perform a template-based analysis using as backgrounds: fluxes from individual sources from 4FGL-DR4 \cite{Ballet:2023qzs}, five IEM components, the isotropic emission component, the Fermi Bubbles template, and Loop I+Sun+Moon component. The five IEM components stand for the bremsstrahlung component, the $\pi^0$ decay production and the IC scattering contribution (itself divided into the Cosmic Microwave Background (CMB), starlight and infrared), and are obtained from \cite{Fermi-LAT:2017opo}.

\begin{table}[h!]
\centering
\caption{Description of the main parameters for the template analysis. In the future we will test the robustness of our results against different values for the energy range, ROI and binning.}
\includegraphics[width=0.7\linewidth]{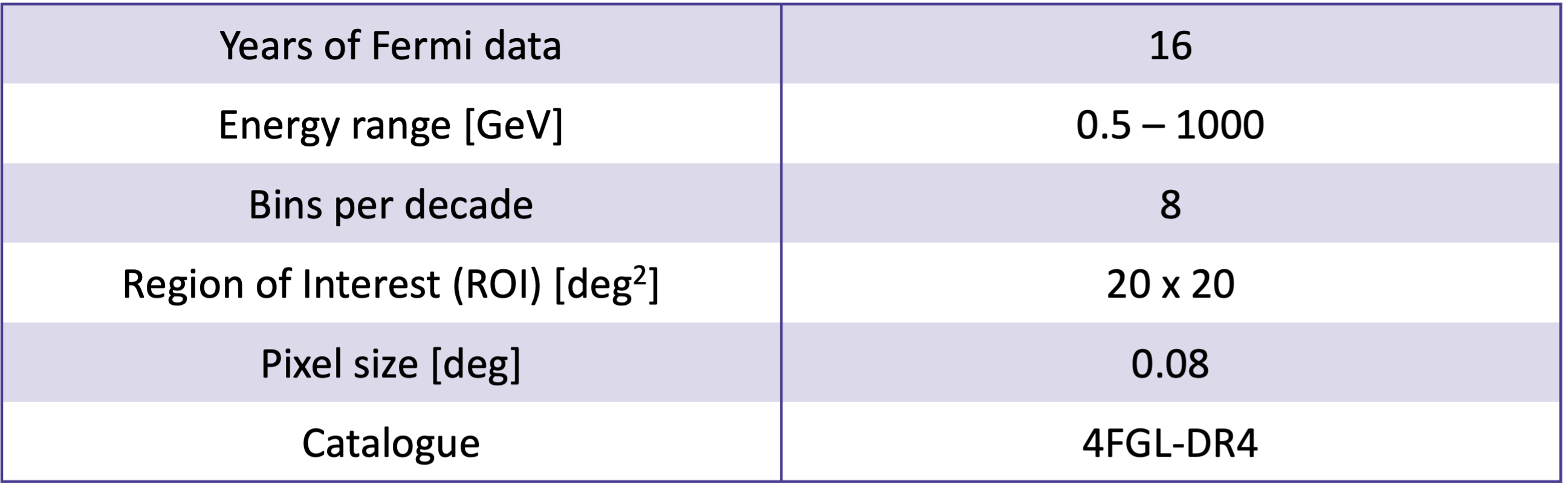}
\label{tab.1}
\end{table}

Our analysis pipeline is based on {\tt FermiPy} \citep{2017ICRC...35..824W}\footnote{See \url{http://fermipy.readthedocs.io/en/latest/}.}. The analysis applied in this work is identical to the one used in our previous work \cite{DiMauro:2023qat}, so we refer to it for further details. In summary, we perform a template-fitting analysis using the likelihood ratio test for each individual object where the main fitting parameter is the normalization of the CR template (i.e. $X_{500}$). Then, we combine the results by summing together the likelihood profiles independently for each energy bin. The significance of the CR hypothesis ($\mathcal{L}(\mu)$) is then tested against the null hypothesis ($\mathcal{L}_{\rm{null}}$):
\begin{equation}
\label{eq:TS}
TS=2~\Delta\mathrm{log}(\mathcal{L})=2~\mathrm{log}\left[\frac{\mathcal{L}\left(\mu\right)}{\mathcal{L}_{\rm{null}}}\right].
\end{equation}
We can assume that the $TS$ distribution follows the $\chi^2$ distribution for two degrees of freedom divided by 2 (Chernoff's theorem \citep{10.1214/aoms/1177728725}). Therefore, a discovery ($5\sigma$) would, in principle, be given by $TS\sim25$. However, due the clusters extensions, we will see in Section~\ref{section:Coma} that the null hypothesis has a broader distribution and as a result, the $5\sigma$ significance will be found for a $TS$ higher than 25.

\subsection{Combined analysis results}
\label{subsec:combined}

Before applying the described analysis pipeline to the real {\it Fermi}-LAT data, we perform recovery tests on mock data to evaluate its performance. We inject the CR emission in the simulations and quantify how precisely we can recover its normalization. We are able to recover the normalization of the injected signals for $X_{500}>0.5\%$ . Moreover, we learn that {\it Fermi} has discovery potential up to as low as $X_{500}= 0.5\%-1\%$ at $TS\sim 10$, increasing up to $TS\sim 200$ for $X_{500}=5\%$.

The first result we quote is the individual and combined $TS$ for the baseline model (i.e. $X_{500}=1\%,\;\alpha_{\rm CR_{p}} = 2.3,\; \eta_{\rm CR_{p}} = 0.75.$), obtaining the highest combined $TS=38$ at $X_{500}\sim 1\%$. For the individual analysis, Coma cluster clearly stands out from the rest, with its highest $TS=30$ for $X_{500}\sim 3\%$. The rest of the individual clusters have $TS$ peaks lower that 25, thus not significant. 

\begin{figure}[h!]
\centering
\includegraphics[width=\linewidth]{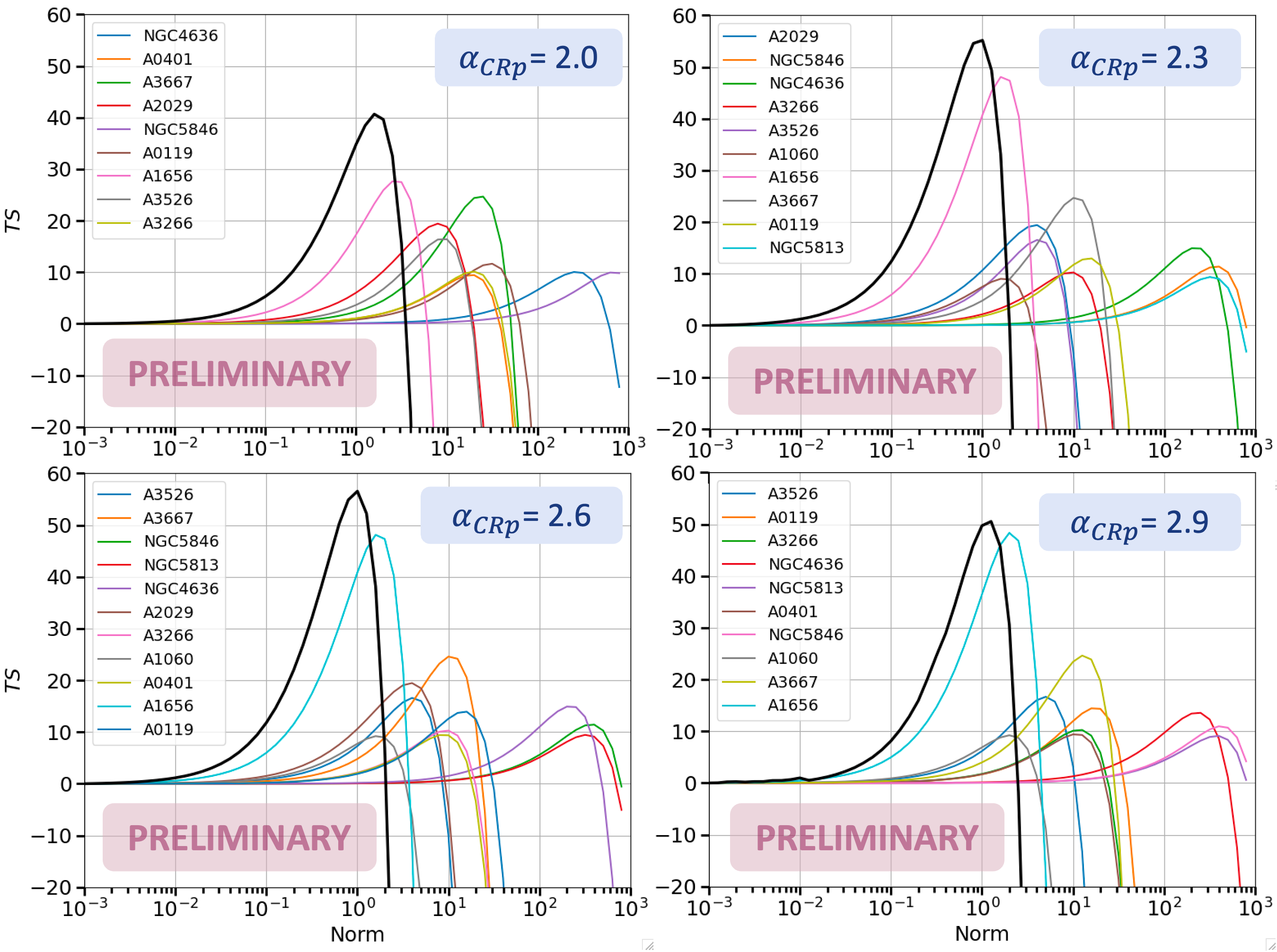}
\caption{$TS$ as a function of the CR normalization ($X_{500}$) obtained from the analysis of individual clusters (colored lines) and combined (solid black). We show only objects for which we obtain $TS>9$. We show, from top left to bottom right, the cases for the $\alpha_{\rm CR_{p}}=2.0, 2.3, 2.6, 2.9$ benchmark models. We clarify that the colors for the individual clusters are not uniquely assigned for the different models.}
\label{fig.2}
\end{figure}

We further explore these results through the analysis using the benchmark models defined in Section~\ref{section:modelling}. The results of the combined and individual $TS$ versus the CR normalization are shown in the panels of Figure~\ref{fig.2}. In this Figure we see that the highest $TS$ peak is reached for the model with $\alpha_{\rm CR_{p}}=2.6$, with $TS=56$, being very close to $TS=54$ reached for the model with $\alpha_{\rm CR_{p}}=2.3$, both of them at $X_{500}\sim 1\%$. For the four benchmark models, the individual $TS$ of Coma also seems to dominate the combined analysis, reaching the highest $TS$ peak for the model with $\alpha_{\rm CR_{p}}=2.6$, at $TS=47$ for $X_{500}\sim 2\%$. None of the rest of the $TS$ values of the individual analysis are significant. 

\begin{figure}[h!]
\centering
\includegraphics[width=0.5\linewidth]{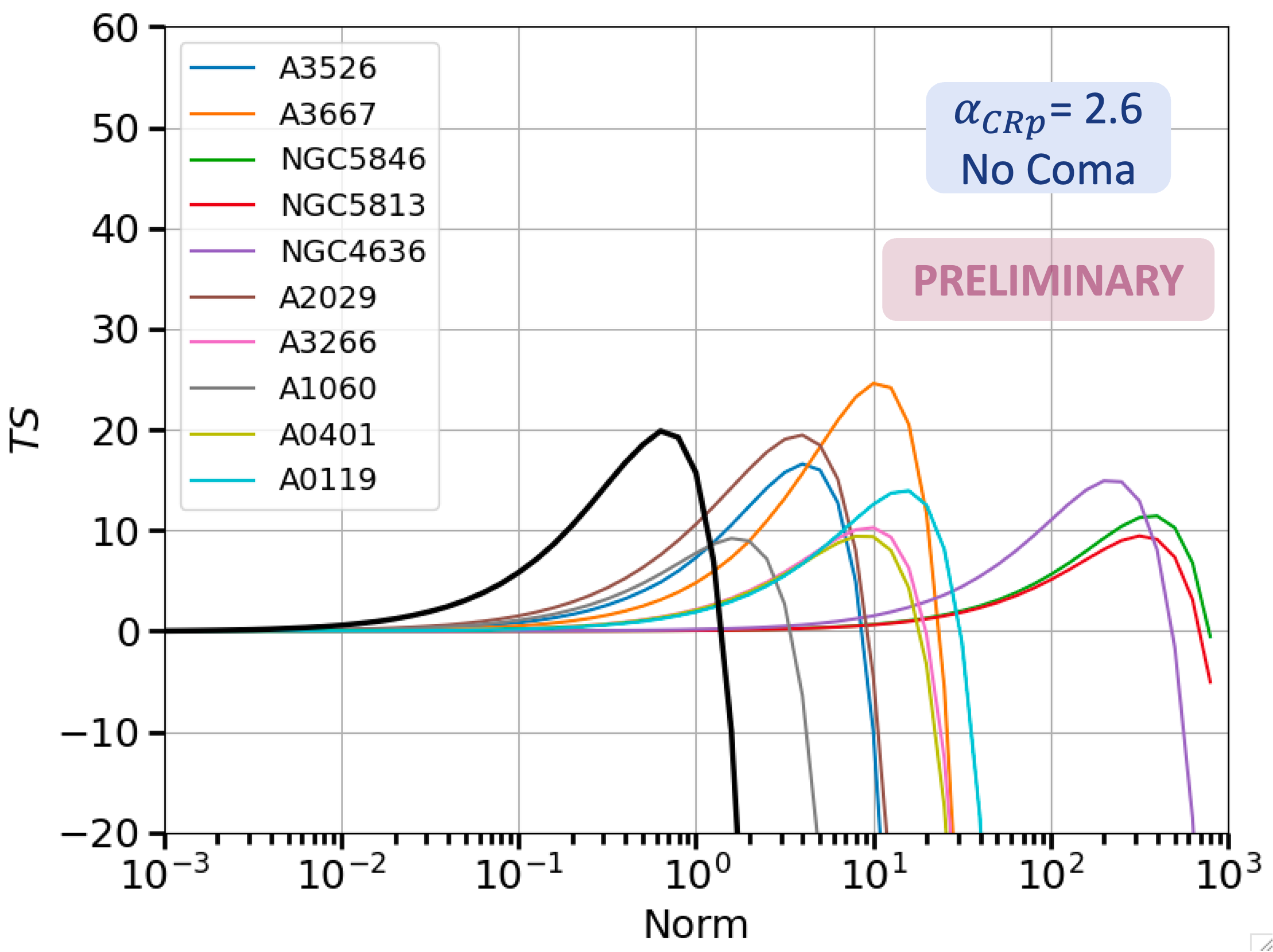}
\includegraphics[width=0.49\linewidth]{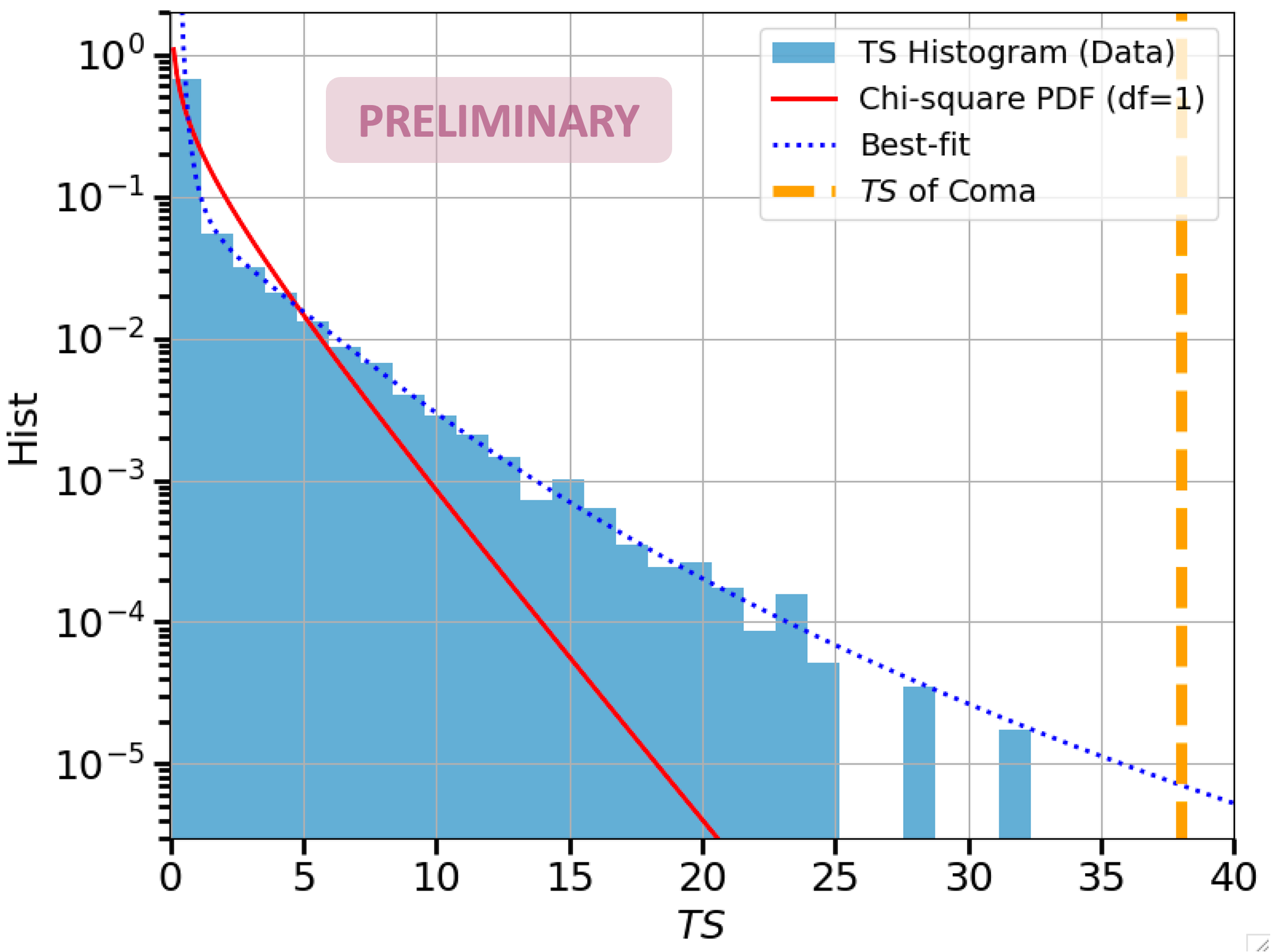}
\caption{{\bf Left panel:} $TS$ as a function of the CR normalization ($X_{500}$) for individual clusters (colored lines) and combined (solid black), assuming the benchmark model with $\alpha_{\rm CR_{p}}=2.6$, excluding the Coma cluster from the original sample. {\bf Right panel:} Normalized histogram of the $TS$ distribution in random directions (blue histogram), $\chi^2$ distribution (red line) for 1 degree of freedom ($X_{500}$) and the best fit (dotted blue line). The vertical yellow dashed line represents the peak of the $TS$ for the Coma cluster using the baseline model.}
\label{fig.3}
\end{figure}

These results suggest a detection in all of the studied cases. Yet, since the individual $TS$ for Coma dominates the combined analysis, we perform an additional test for the best detection case ($\alpha_{\rm CR_{p}}=2.6$) without including Coma. These results are shown in the left panel of Figure~\ref{fig.3}. This plot showcases that, when Coma is removed from the analysis, the combined $TS$ peak drops until $TS=20$ at $X_{500}\sim 7\%$, proving that the results obtained from the whole sample are clearly dominated by this object. In light of these results, this new value for the $TS$ is below the detection significance, thus we will translate these results into constraints in the $X_{500}-\alpha_{\rm CR_{p}}$ parameter space.

\subsection{Coma cluster's case}
\label{section:Coma}

As discussed in Section~\ref{sec:analysis}, in the case where we have perfect knowledge of the backgrounds, we can establish that a discovery of $5\sigma$ should correspond to $TS \sim 25$. Nevertheless, the analysis of real data for extended sources deviates significantly from this asymptotic scenario. For our case, to convert the $TS$ for a CR signal into a significance, we need to build the $TS$ distribution using random blank sky directions. The methodology followed is very similar to the ones we used in our previous work \cite{DiMauro:2023qat}, with the difference that we increased the number of random directions used to 50000. The distribution of the $TS$ for the baseline model is shown in the right panel of Figure~\ref{fig.3}. There is a prominent tail at larger $TS$ values compared to that of the $TS$ distribution following. The first thing we notice is that, as expected, there is a prominent tail at larger $TS$ values compared to that of the $TS$ distribution following the $\chi^2$ one. The yellow line in this plot represents the $TS$ peak for Coma for the analysis using the baseline model ($TS=37$). Thus, according to the best fit of this histogram, this $TS$ value is associated with a local $p$-value of $7\times 10^{-6}$ and a significance of 4.3$\sigma$.

There exist previous claims in the literature for the detection of extended emission from the Coma cluster around $TS\sim 20-50$ \citep{Xi:2017uzz,AdamEtAl2021,Baghmanyan:2021jwg}. Despite the fact that these papers assume specific models for the hadronic emission of photons or simple geometrical extended templates, our results seem to be in agreement. The Coma cluster is also a very complex region: in the radio band it shows an extended, non-thermal, radio halo at the core and a diffuse relic at the virial radius; in X-rays it shows a sub-group linked to the AGN NGC~4839. There are also some claims that the 4FGL J1256.9+2736 source from the 4FGL-DR4 catalogue \cite{Ballet:2023qzs} could correspond to NGC~4839. With this complex picture, in the future we will further scrutinize the possible origin of the 4.3$\sigma$ detection in the Coma region, with different set-ups including the aforementioned AGN and different simple spatial models for the CRs diffuse emission.

%\begin{table}[h!]
%\centering
%\caption{Electron number density (left panel) and electron pressure (right panel) profiles obtained for the Virgo cluster using the parametrized models from \cite{Ghirardini:2018byi} and \cite{2010A&A...517A..92A}, respectively (solid black line). These parametrizations are compared to ACCEPT data \cite{2009ApJS..182...12C} (red dots). For most of the clusters in our sample with ACCEPT data available, the generalized models yield flatter profiles compared to the fitted profiles (dashed green line), being more conservative in terms of $\gamma$-ray emission.}
%\includegraphics[width=\linewidth]{ICRC2025_template/table_coma_models.png}
%\label{tab.2}
%\end{table}

\bibliographystyle{JHEP}
{\footnotesize\bibliography{paper}}

\end{document}